\documentclass[12pt]{article}
\usepackage{amsmath,amssymb}

\newcommand\be{\begin{equation}}
\newcommand\ee{\end{equation}}
\newcommand\bea{\begin{eqnarray}}
\newcommand\eea{\end{eqnarray}}
\newcommand\nn{\nonumber}
\newcommand\ba{\(\begin{array}}
\newcommand\ea{\end{array}\)}

 \def\p{\partial}
 
 \def\a{\alpha}
 \def\b{\beta}

 \def\e{\epsilon}

 \def\l{\lambda}

 \def\t{\tau}
 
 \def\z{\zeta }

 \newcommand{\tr}{\text{tr }}

\def\ello{\ell^{(1)}}
\def\ellt{\ell^{(2)}}
\def\zo{z^{(1)}}
\def\zt{z^{(2)}}
\def\zeo{\z^{(1)}}
\def\zet{\z^{(2)}}

\def\so{s^{(1)}}
\def\st{s^{(2)}}

\makeatletter
\@addtoreset{equation}{section}
\makeatother

\newcommand{\cref}[1]{(\ref{#1})}

\allowdisplaybreaks
\begin{document}

\setlength{\oddsidemargin}{0cm}
\setlength{\baselineskip}{7mm}

\setlength{\oddsidemargin}{0cm}
\setlength{\baselineskip}{7mm}

\begin{titlepage}

\renewcommand{\thefootnote}{\fnsymbol{footnote}}

\vspace*{0cm}
    \begin{Large}
    \begin{bf}
       \begin{center}
         {Annulus amplitude of FZZT branes revisited}
       \end{center}
    \end{bf}
    \end{Large}
\vspace{0.7cm}

\begin{center}
Jae-Hyuk O{\sc h$^{1}$}\footnote{e-mail address :
jack.jaehyuk.oh@gmail.com}, Jaemo P{\sc ark$^{2}$}\footnote{
e-mail address : jaemo@postec.ac.kr }
 and Chaiho R{\sc im$^{3}$}\footnote
           {
e-mail address :
rimpine@sogang.ac.kr}\\

\end{center}

\vspace{0.7cm} \centerline{${}^1$\sl\small Harish-Chandra Research
Institute, 
Allahabad-211019, India }

\centerline{${}^2$\sl\small Physics Department
and PCTP,
Postech, Pohang,
790-784, Korea}
\centerline{${}^3$\sl\small Dept. of Physics and CQUeST,
Sogang University, Seoul 121-742, Korea} \vspace{0.7cm}

\begin{abstract}
\noindent
We revisit the annulus amplitude of FZZT branes
with general matter sectors $(r,s)$ using the recent
development of matrix model and minimal Liouville gravity.
Following the boundary description of the 1-matrix model
and bulk resonance transformation between primary operators
we find the consistency of the brane decomposition into $(1,1)$-branes.
We also investigate the corresponding results
obtained directly from the minimal Liouville gravity
and demonstrate the perfect agreement with the matrix results.
\end{abstract}
\vfill

\end{titlepage}


\setcounter{footnote}{0}


\section{Introduction}
The  quantum gravity in two space-time dimensions
can be described in terms of Liouville gravity
\cite{polyakov}
and its non-perturbative  effect of
interaction with matter
is reliably investigated
if conformal symmetry is maintained.
The interaction with minimal matter is studied in the name of
minimal Liouville gravity (MLG).
The minimal gravity is simple because
the number of primary fields is finite and
the exact correlation numbers (expectation values of
coordinate-integrated form of correlation functions)
can be obtained.

MLG  is also studied in the context of matrix models.
$(2, 2p+1) $ minimal Liouville gravity
is described by the hermitian 1-matrix model
(sometimes called as $p$-critical model) \cite{1-matrix}
and by 2-matrix model \cite{2-matrix}
the  minimal $(q_1,q_2)$   unitary theory  (with $q_1 < q_2$ co-prime).
The comparison of the matrix model with MLG is a non-trivial task  \cite{MSS}.
Nonetheless, the parameter dependence of MLG
is conjectured  on the fluctuation sphere and its exact form is provided
for the case of the Lee-Yang matter ($p=2$) \cite{alzam}.
For $p \ge 3$ one needs to consider  the resonance between primary operators.
The exact bulk resonance transformation
(BZ transformation) is conjectured for the  $p$-critical case   \cite{BZ}
and is tested   up to some of five-point correlations \cite{BT}.

When one considers  boundaries in MLG, one needs to specify the
boundary condition, which is represented by D-branes.
Possible  D-branes,  FZZT in MLG  is discussed in \cite{FZZ,T}.
The boundary state is given by the tensor product of that of
Liouville theory and that of minimal model
and is specified by the continuous boundary  parameter  $s$
and  by the two integers $(k \leq q_1, \ell \leq q_2)$
in the $(q_1,q_2)$ MLG.
It is conjectured that not all of these states are independent
but is argued that general boundary states
coming from $(k, \ell) $ states are linear combination of $(1,1)$-brane.
Specifically,
\begin{equation}
|s; (k,l)\rangle
=\sum_{m'=-(k-1),2}^{k-1}~
\sum_{n'=-(l-1),2}^{l-1}|s+im'\frac{1}{b}+in'b;1,1\rangle
\label{branerelation}
\end{equation}
where $b=\sqrt{\frac{q_1}{q_2}}$.
This relation is checked at the
ground ring level in \cite{SS,BM}.
With this conjecture,
most of the interest is  centered on (1,1) brane
whose matrix object is associated with the macroscopic loop operator
\begin{equation}
\left\langle
\tr \frac{1}{u_0 z-M}
\right\rangle
\end{equation}
where $u_0$ is proportional to the square root of the bulk
cosmological constant with KP scaling 1
(We do not elaborate on this fine tuning at the critical limit further;
one may refer to {\it e.\/g.\/} \cite{B-1-matrix, B-2-matrix})
and  $z$ is related to the continuous boundary cosmological constant
parameter $s$
\begin{equation}
z= \cosh ( \pi b s)\,.
\end{equation}
The disk partition function of the matrix model is given as
${\cal Z}= \langle \tr \ln (z-M)  \rangle$.

For other branes with general matter sector
little study has been done until recently.
Indeed,  the disk partition function of $p$-critical model with BC $(s, (1,m))$  is  given as
\cite{B-1-matrix}
\begin{equation}
{\cal  Z}_{\rm disk} (s; (1, m)) = \langle \tr \log F_m(z, M ) \rangle
\,,~~ F_m(z, M) = \prod_{k=-(m-1):2}^{m-1} (u_0 z_k - M)
\label{matrixrelation}
\end{equation}
where $z_k =  \cosh (\pi b s_k)$ with $s_k = s+ i b k $. This
proposal is obviously consistent with Eq.~(\ref{branerelation}) with
$k=1$ since we are dealing with $(2, 2p+1)$ case out of general
$(q_1,q_2)$:
The partion function is simply addition of that of $(1,1)$-boundary
with cosmological constant parameter shifted by
suitable imaginary value,
consistent with the brane decomposition of MLG.
In addition, it is obvious how to generalize the above
proposal  Eq.~(\ref{matrixrelation}) to  $(q_1,q_2)$-model
by considering 2-matrix model
\cite{B-2-matrix}.
Nontrivial tests for this proposal were carried out in
\cite{B-1-matrix, B-2-matrix} at the disk level.
It is confirmed that the disk one and two-point correlations
in the matrix model  reproduce the known
results of Liouville theory\cite{FZZ}.

This idea of decomposition of
the branes are very intuitive and the idea should go beyond the disk
boundary.
Given the prescription of  Eq.~(\ref{matrixrelation}),
it is straightforward to work out the corresponding annulus amplitude in
the matrix model and compare the results of the matrix model
with the corresponding MLG,
which is the main theme of this paper.  In this paper we carefully work
out the annulus amplitude and find the perfect agreement between
the matrix model proposal and the MLG computation, thereby
confirming the proposal of \cite{B-1-matrix, B-2-matrix} at annulus
geometry.

Incidentally, this solves the confusions recently
raised on the MLG results \cite{kutasov, aw}.
The annulus amplitude is evaluated in \cite{martinec}
using the boundary Louville field theory and lattice height model of
$A_{q_1-1}$ series \cite{saleur}. For example, for the $ (1,1)$-boundary,
the annulus amplitude is given as
\be
\label{annulus-Liouville}
{\cal Z}(s,1|s', 1) =
 \int_{-\infty}^\infty  \frac{d \nu}\nu
\frac{ \cos (\pi s \nu)  \cos (\pi s' \nu)  \sinh(\pi (q_1-1) \nu/b
)} {\sinh (\pi \nu/b)~ \sinh(\pi q_1 \nu/b) }\,.
\ee
This shows a subtle point  since the amplitude is to be
regulated to avoid the divergence at $\nu=0$. The subtlety raises
questions about the universal form of the annulus amplitude
\cite{kutasov, aw} and about  the decomposition  into $(1,1)$-branes \cite{aw}.

The content of the paper is as follows. In section \ref{section2},
we check the annulus amplitude in the matrix model. For  $ (1,1) $
boundary,  we use the boundary 1-matrix model \cite{B-1-matrix}
and evaluate the amplitude using the filling fraction
representation. In this way, the universal contribution of the
amplitude is identified.
And it is straight-forward to write down the
annulus amplitude for branes with general matter sectors. In section
\ref{section3}, we revisit the Liouville theory computation of
the minimal gravity obtained in \cite{aw}. After
using the summation formula to get the compact form of the
amplitude, one can explicitly demonstrate that the result reduces to
the formula (\ref{annulus-Liouville}).
In addition, we present the annulus
amplitude for the general boundaries  and
find the result consistent with the brane-decompostion.
Section \ref{section-sum} is the conclusion and discussion where
bulk correlation  in the annulus is presented
for the $p$-critical model using the
BZ transformation. In appendix, one can find detailed
calculations.

\section{Annulus amplitude in the matrix model}
\label{section2}

The $p$-critical model ($q_1=2, q_2= 2p+1$)  is described in terms
of one-matrix model. Even though the $p$-critical model  is
non-unitary series, the finite  number of primary operators produces
many properties sharing with the unitary series ($q_1 \ge 3$). Thus,
we start with the one-matrix model for simplicity. The annulus
amplitude is obtained from the two-loop correlation
\be \label{annulus-matrix} W_2
(z^{(1)}, z^{(2)} ) = \left\langle \tr \left( \frac1{u_0 z^{(1)} -
M}\right) ~ \tr  \left(\frac1{u_0 z^{(2)}-M}\right) \right\rangle
\ee
where $z^{(i)} = \cosh( \pi bs^{(i)})$.
Explicit evaluation shows  \cite{amb,DKK, eynard}
\be W_2
(z^{(1)}, z^{(2)} ) = \frac{\partial}{\partial z^{(1)}}
\frac{\partial}{\partial z^{(2)}} \log \left( \frac{\z^{(1)} -
\z^{(2)} }{z ^{(1)} -z^{(2)}}\ \right) =-  \frac{\partial}{\partial
z^{(1)}} \frac{\partial}{\partial z^{(2)}} \log (\z^{(1)} + \z^{(2)}
) \ee
where  $\z^{(i)} = \cosh( \pi bs^{(i)}/2)$ covers  the
double-sheet parameter space. The definition of the annulus
amplitude $ W(\zo, \zt)=\frac{\partial}{\partial \zo}
\frac{\partial}{\partial \zt} {\cal Z}( \zo, \zt) $ results in \be
\label{annulus-amplitude} {\cal Z}(\zo, \zt) = -  \log (\z^{(1)} + \z^{(2)}
) + f_1(\zo) + f_2(\zt) \ee
where $f_i $'s are function of
$\zo$ or $\zt$ only. The result is consistent with Eq.~(\ref{annulus-Liouville}).

The remaining subtle point is the
regularization dependency and the universal behavior of the annulus
amplitude \cite{kutasov, aw}.
To clarify these, we provide another useful and simple formula
for the annulus amplitude in terms of filling fraction  integral
representation and BZ resonance in \cite{BZ}. The nice
feature of this representation is that one can pin-point  the
universal contribution precisely. We may put  the two-loop
correlation (\ref{annulus-matrix}) using the Laplace transformation
\be \label{2-loop} W(\zo, \zt)
= \int_{\ello, \ellt  \ge 0}
e^{-(\ello \zo + \ellt \zt )} C(\ello, \ellt)
\ee
where $C(\ello, \ellt) $ is interpreted as the annulus amplitude with fixed lengths
\be
\label{2-loop-length} C(\ello, \ellt) = \int_1^{\infty } dx
\int^1_{0} d\tilde y \langle x | e^{ \ello( u_0 d^2 -u) }| \tilde y
\rangle \langle \tilde y | e^{ \ellt( u_0 d^2 -u) }| x \rangle
\ee
where $d$ denotes the differential operator with respect to the
filling fraction $x, \tilde y$  of the matrix eigenvalues.
Note that the integration range does not overlap
except $x=\tilde y \ne 0$ so that  $ \langle \tilde y| x\rangle =0$.
To proceed, we change the variables $x \to 1-x$ and
$\tilde y \to 1 - \tilde y$ and use  $ x +Q_p (u) =0$
which sets  $u$ as  a certain function of $x$
through BZ transformation.
(Note that the string equation is given as $Q_p(u_*)=0$). This
identification translates the matrix result (kdV frame) into the
field theory one  (CFT frame). $Q_p (u)$ is given in terms of  the
Lengendre polynomial  $L_p( \xi)$ with $\xi = u/u_0$ \cite{BZ}
(in the absence of the bulk couplings),
\be Q_p(u) =  \frac{L_{p+1}(\xi ) - L_{p-1}(
\xi )}{2p+1}\,.
\ee

One may evaluate (\ref{2-loop-length})
with the help of the momentum integration
\be
C(\ello, \ellt)
=
\frac1{\sqrt{\ello  \ellt}}
\int_{-\infty}^{0}  dx \int^{1}_0 d \tilde y
e^{-  (x - \tilde y)^2 (\ell_1 + \ell_2)/(\ello \ellt)}
e^{-\ello u(x) -\ellt u( \tilde y ) }\,.
\ee
$C(\ello, \ellt)$ is
proportional to $\sqrt{\ello \ellt} /(\ello + \ellt) $
as $\ello, \ellt \to 0$.
After integration over the length variables of the annulus amplitude (\ref{2-loop})
one has  the universal form
\be
\label{annulus-integral}
{\cal F}(\zo, \zt) =
 \int_{x, \tilde y} \frac{ e^{- |x - \tilde y|R_0} }{(x-\tilde y)^2}
+ f_1 (\zo) + f_2( \zt) \ee where $\int_{x, \tilde y} \equiv
\int_{-\infty}^{0}  dx \int^{\infty}_0 d \tilde y $ and $
R_0=  \sqrt2 (\zeo +\zet) $.
We distinguish
${\cal F}(\zo, \zt)$  from  ${\cal Z}( \zo, \zt) $ for later use. In
addition, we change the integration limit of $\tilde y$ from 1 to
$\infty$ since this addition does not change the universal part
because the  universal contribution comes from
the region where the string equation $Q_p(u_*) =0$ is satisfied (at
$x=\tilde y=0$), whose solution is $u_* = u_0$
(See details in Appendix A).

It is noted that the integral is not convergent at $x= \tilde y=0$.
To make the integration finite, one may choose the
integration constants $f_i $'s  so
that the integral ${\cal F}(0, 0)=0$
\be \label{annulus-partition_final}
{\cal F}(\zo, \zt) = -\log
\left( \frac{\zeo + \zet }2 \right)
\ee which is consistent with Eq.~(\ref{annulus-amplitude}).
One may wonder if one can
remove $\log( \zo -\zt) $ by a suitable regularization. However, it
is obvious that that choice is impossible.

The annulus amplitude with boundaries
$ (\so, (1,m)) ,  (\st ,(1,\ell) ) $
 is  proposed in  \cite{B-1-matrix}
\be 
\label{z-def}
{\cal Z}_{\rm ann}(\so,(1, m)| \st, (1,\ell))
 = \langle \tr \log F_m(\zo, M )~ \tr \log F_\ell( \zt, M ) \rangle_c
\ee
where $\langle~\rangle_c$ stands for the connected part of the
partition function. According to this,  the amplitude is
consistent with the decomposition of the $(1,1)$-branes
\cite{kutasov}
\be \label{Zlm}
{\cal Z}_{\rm ann}(\so,(1, m)| \st, (1,\ell)) 
\! =\sum_{m'=-(m-1);2}^{m-1}\sum_{\ell'=-(\ell-1);2}^{\ell-1}
{\cal Z}_{\rm ann} (\so_{m'}, (1,1) |\st_{\ell'},(1,1))\,. \ee
Similar decomposition for  the boundary  2-matrix model can be checked 
\cite{B-2-matrix}
by extending (\ref{z-def}) into 2-matrix version
\begin{align}
\label{two-matrix-decomp}
&{\cal Z}_{\rm ann}(\so, (r,m)| \st, (s, \ell))
\nn \\
&~~~~~~~
 =
\sum_{s'=-(s-1);2}^{s-1}\sum_{m'=-(m-1);2}^{m-1}
\sum_{r'=-(r-1);2}^{r-1}
\sum_{\ell'=-(\ell-1);2}^{\ell-1}
{\cal Z}_{\rm ann} (\so_{s',m'},(1,1)|\st_{r',\ell'},(1,1))
\end{align}
where $s_{r,m}^{(i)}= s^{(i)}+ i r/b + im b$.

\section{Annulus amplitude in minimal Liouville gravity}
\label{section3}

Now let us investigate  $(q_1, q_2)$-MLG: The
annulus amplitude is considered
in \cite{aw, Ito}.
For  $(1,1)$-boundary  one has\footnote{
Normalization of the Liouville part is taken so that $2\sqrt2 \pi^2$ 
is absent and - sign correction is done in (3.10) of \cite{aw}}
\be \label{Zq1q2}
{\cal Z}((1,1),s|(1,1),s') = - \frac{1}{ 2 q_1
q_2} \int_{-\infty}^\infty \frac{d\eta}\eta \frac{\cos (\sqrt{q_1 q_2}
\eta s) \cos (\sqrt{q_1 q_2}  \eta s' ) \sinh \eta } {\sinh (q_1 \eta)
\sinh (q_2 \eta)} F_{1,1}(i \eta)
\ee
where \be \label{F11-sum}
F_{1,1} (z) = \sum_{\a =1}^{q_1-1} \sum_{\b =-(q_2-1)}^{q_2-1}
\frac{\sin (\pi t /q_1 ) \sin (\pi t /q_2 )} {\cos (\pi t/(q_1 q_2))
-\cos z}
\ee with $t= \a q_2 + \b q_1 $.
After summation,
$F_{1,1}(z) $  is given
in a compact form\footnote{One can use the same
trick given in \cite{kutasov} using the pole structure and the large
imaginary behavior in $z$ for $q_1 < q_2$.
See the details in App.~C.}
\be \label{F11} F_{1,1}
(z) =- 2 q_1 q_2 \frac{\sin (z (q_1-1) q_2 ) \sin (z q_1 )} {\sin(z
q_1 q_2) \sin(z)}\,. \ee
This shows that the annulus amplitude
reproduces\footnote{It seems that (3.16) in
\cite{aw} does not go with this observation.} exactly the same result
(\ref{annulus-Liouville})  ($\eta \to \pi \nu /\sqrt{q_1q_2}$).
Thus, one  concludes that the annulus amplitude for
$(q_1, q_2)$-minimal gravity will be in the form  \cite{kutasov}
\be
{\cal Z}(\zo, \zt)
 = \log \left( \frac{\zeo_{q_1q_2} - \zet_{q_1 q_2} }
{T_{q_1} (\zeo_{q_1 q_2}) - T_{q_1} (\zet_{q_1q_2}) } \right)
\ee
where $\zeta^{(i)}_{q_1q_2} = \cosh(\pi s^{(i)}/\sqrt{q_1 q_2})$.

To see the general boundary amplitude, let consider $q_1 =2$ case
first.  According to the result of \cite{aw}, the numerator  term
$\sin (\pi t /q_2 )$ in (\ref{F11-sum}) is modified into
\begin{align}
F_{m,\ell} (z)
&= \sum_{\b =-(q_2-1)}^{q_2-1}
\frac{\sin (\pi t /q_1 )
\Big\{ \sin (\pi t m  /q_2 ) \sin (\pi t \ell  /q_2 )/ \sin (\pi t   /q_2 )\Big \} }
{\cos (\pi t/(q_1 q_2)) -\cos z}
\nn\\
&= - 2 q_1 q_2 \frac{\sin (z (q_1-1) q_2 )  \sin (z q_1 \ell)}
{\sin(z q_1 q_2) \sin(z)}\,.
\end{align}
where $t = q_2 + \b q_1$ and without loss of generality $1 \le \ell,
m \le (q_2 -1)/2$. Putting this into the annulus amplitude
(\ref{Zq1q2}) one has the matrix  result (\ref{Zlm})
(Note that the numerator
$\cos (\sqrt{q_1 q_2} \eta s) \sinh  (\eta q_1 m )/
\sinh(\eta q_1) $ is decomposed into the sum of
$\cos (\sqrt{q_1 q_2} ~ \eta s_{m'}) $ ).

It is a simple matter to confirm the unitary series $q_1 \ge 3$,
using the formula (\ref{sum3}) and (\ref{sum4}) in App.~C
that the amplitude indeed satisfies the decomposition
(\ref{two-matrix-decomp})

\section{Conclusion and discussion}
\label{section-sum}

We provide the explicit form of the annulus amplitude for $(1,1)$ boundary
in two different approaches,
one using the boundary matrix model and
the other using the  minimal Liouville gravity.
The universal part of the matrix model agrees with the one
given from the Liouville gravity side,
even though one needs to  regularize the annulus amplitude.
It is noted that  the universal contribution
of the annulus amplitude
of the 1-matrix model
(the $p$-critical model, $q_1=2$ and $q_2 = 2p+1$)
is given as
$\log (\z^{(1)} + \z^{(2)} )$
rather than $\log(\z^{(1)} - \z^{(2)} )$ \cite{kutasov}.
On the other hand, the annulus amplitude
of the general boundary
is decomposed into the sum of $(1,1)$ boundaries
as proposed in \cite{kutasov},
with the Liouville boundary parameters are (imaginary) shifted.

After this convincing evidence for the annulus amplitude
from the matrix side,
one may calculate  bulk correlation of the $p$-critical model in the annulus
using the formula in (\ref{annulus-amplitude})
if one applies the BZ transformation 
in the presence of the bulk source
\be
Q_p(u, \l_k) =
 \frac{L_{p+1}(\xi) - L_{p-1}(\xi)}{2p+1}
+ \lambda_{k} L_{p-k} + O(\lambda^2)
\ee
where $\lambda_{k}$ is the source to the dressed bulk operator
$O_k = \int_M e^{2 b \a_k \varphi} \Phi_k$
with $\a_k = (k+1)/2$  ($k=2, \cdots, p$).
$\varphi$ is the Liouville field
and $\Phi_k$ represents
the matter field with $\Phi_1=I$.
The bulk correlation is defined as 
${\cal F} (\zo, \zt; O_k)
\equiv - \left. \frac{\p}{\p \l_{k}} 
{\cal F}(\zo, \zt)\right|_{\l_k =0}$.
Using the properties; $x(\xi, \l_k) = - Q_p(u, \l_k)$
so that $x \equiv x(\xi) = -Q_p(u,0)$ 
with the conditions $x(\xi=0) =0$ and 
\begin{align}
\left.
\frac{\p dx(\xi, \l_k)}{\p \l_k}\right|_{\l=0}
&= dx ~\frac{L'_{p-k}(\xi) }{L_p(\xi)}
\,,~~~~
\left.
\frac{\p x(\xi, \l_k) }{\p \l_k}\right|_{\l=0} = - L_{p-k}(\xi)
\end{align}
one has finite result  with KP scaling factor $u_0^{-2\a_k}$
(noting that the subtracted term in 
(\ref{annulus-integral})
or (\ref{full})
has no $\l_k$-dependence)
\begin{align}
\label{bulk-annulus}
{\cal F} (\zo, \zt; O_k)
= - \int_{x, \tilde y}
\frac{ e^{- |x - \tilde y|R_0 } }{(x-\tilde y)^2}~
g(x, \tilde y)\,, 
\end{align}
where $g=g_e + g_o$
\begin{align}
\label{g-function}
g_e(x, \tilde y)
&=g_e(\tilde y, x)
=\frac{L'_{p-k}(\xi) }{L_p(\xi) } +
\frac{L'_{p-k}(\tilde\xi)} {L_p(\tilde \xi) } +
2 \frac{ L_{p-k}(\tilde \xi) - L_{p-k}(\xi)}{\tilde y -x}
\nn\\
g_o(x, \tilde y)
&= -g_o(\tilde y, x)
=(L_{p-k}(\tilde \xi)  - L_{p-k}(\xi))R_0 \,.
\end{align}
  
The integration is not simple to carry out. 
By noting that rescaling $x$ and $\tilde y$ by $1/R_0$ is broken in $g$, 
one may evaluate ${\cal F} (\zo, \zt; O_k)$
in $1/R_0$ expansion. In fact, the $x$ and $\tilde y$ symmetry 
enforces the odd power of $1/R_0$ to vanish
and has the form of expansion  
$ F (\zo, \zt; O_k)  =  \sum f_n R_0^{-2n}$
where  $f_n$ is a constant which depends only on $p$ and $k$.
Explicit calculation shows that 
$f_0 = (p+1-k)(p-k)/2$ and $f_2=0$ 
(see Appendix C).

\section*{Acknowledgements}

This work is supported
by the National Research Foundation of Korea (NRF)
grant funded by the Korea government (MEST) 2005-0049409 (CR and JP),
R01-2008-000-20370-0 and  2009-0085995 (JP),
and 11-R\&D-HRI-5.02-0304 through Harish-Chandra Research Institute in India 
(JO). JP thanks APCTP for its stimulating environment for research.
JO thanks CQUeST(Sogang University) for hospitality
where this work started during his visit
and J.-E.  Bourgine for useful discussion. 

 \section*{Appendix}
\appendix

In this Appendix, we provide details of the calculation
needed in the text.

\section{Evaluation of  the annulus amplitude}

To find  ${\cal F}(\zo, \zt)$ in
(\ref{annulus-integral}),
we first integrate    (\ref{2-loop}) over $\ell^{(i)}$'s
\be
W(\zo, \zt)
=
\int_{-\infty}^{0}  dx \int^{\e}_0 d \tilde y
\int_{p_1, p_2}
\frac{e^{i( p_1-p_2) (x - \tilde y)} }
{(u_0 \, p_1^2  +u(x) + \zo)
(u_0\,  p_2^2 + u( \tilde y ) +  \zt)}\,.
\ee
The annulus amplitude is given as
$ {\cal F}( \zo,  \zt) = {\cal F}_0 (\zo, \zt)  + f_1 (\zo) + f_2(\zt) $
where $f_i$'s are integration constants and
\be
{\cal F}_0 (\zo, \zt) =
\int_{x, \tilde y}
\int_{p_1, p_2}
e^{i( p_1-p_2) (x - \tilde y)}
\log (u_0\, p_1^2  +u(x) + \zo)
\log(u_0\, p_2^2 + u( \tilde y ) + \zt)
\ee
with the shorthand notation\footnote{
The integration range is originally   $0< \tilde y <1$.
We put $\e$ in the integration limit for later convenience. },
$
\int_{x, \tilde y} \equiv
\int_{-\infty}^{0}  dx \int^{\e}_0 d \tilde y$.
After integration by part of the momenta one has
\begin{align}
{\cal F}_0 (\zo, \zt) &=
\int_{x, \tilde y}
\int_{p_1, p_2}
\frac{ \log (u_0\, p_1^2  +u(x) + \zo)
\log(u_0\, p_2^2 + u( \tilde y ) + \zt))}
{(x-\tilde y)^2}
\frac{\partial}{\partial p_1}
\frac{\partial}{\partial p_2}
e^{i( p_1-p_2) (x - \tilde y)}
\nn\\
&=
\int_{x, \tilde y}
\frac1{(x-\tilde y)^2}
\int_{p_1, p_2}
e^{i( p_1-p_2) (x - \tilde y)}
\left(
\frac1{p_1+ i \sqrt{\xi(x) + \zo}}
+
\frac1{p_1- i \sqrt{\xi(x) + \zo}}
\right)
\nn\\
&~~~~~~~~~~~~~~~~~~~~~~~~~~~~~~~~~~~
\times
\left(
\frac1{p_2+ i \sqrt{\xi(\tilde y) + \zt}}
+
\frac1{p_2- i \sqrt{\xi(\tilde y) + \zt}}
\right)
\nn\\
&=
\int_{x, \tilde y}
\frac{
e^{- |x - \tilde y| R_{12}(\xi, \tilde \xi)}
}{(x-\tilde y)^2}
\label{F0}
\end{align}
where $\xi = u(x)/u_0 \ge 1$,  $\tilde \xi = u( \tilde y)/u_0 \le 1 $, and 
$  R_{12}(\xi, \tilde\xi) = \sqrt{\xi + \zo}+ \sqrt{\tilde \xi + \zt}$  is real and positive.

The integration in (\ref{F0}) will give divergent contribution  in general when $x= \tilde y$.
One can make the amplitude
finite after  subtracting this divergence using
the integration constants  $f_i (z^{(i)})$'s,
since the divergence is independent of  $z^{(i)}$'s.
Suppose one requires  $F(\zo=\zt= 0 |R )=0$,
one has
\be
\label{full}
F(\zo, \zt |R )=
\int_{x, \tilde y}
\frac{
e^{(x- \tilde y) R_{12}(\xi, \tilde \xi)}
}{(x-\tilde y)^2}- (  R_{12}(\xi, \tilde \xi) \to  \tilde  R_{12}(\xi, \tilde \xi) )
\ee
where $ \tilde  R_{12}( \xi, \tilde \xi) =   R_{12}(\xi, \tilde \xi)|_{\zo=\zt=0}$.

Let us evaluate the universal contribution of  $F(z_1, z_2 |R )$ in (\ref{full}).
One may add  the contribution $\tilde y >1$ without affecting the universal part.
Note also  that at  $x=\tilde y =0$,
the string equation $Q_p(u)=0$
has the solution $\xi = \tilde \xi = 1 $.
Therefore, the universal contribution
can be of the form
if one put $ R_{12} \to R_0  =\sqrt{1 + \zo}+ \sqrt{1 + \zt}$
and $\epsilon \to \infty$
\be
\partial^{2}_{R_0}F(z_1, z_2 |R_0)
=
\lim_{\epsilon \to \infty}
\int_{x \tilde y} e^{-(x+\tilde{y})R_0}=\frac{1}{R_0^2}\,.
\ee
Integrating over $R_0$ twice, one has
\be
\label{universal}
F(\zo,\zt| R_0)= -\log \left( \frac {R_0(\zo, \zt)} {R_0(0,0)}
\right) =- \log \left(
\frac{\z_1 + \z_2}2 \right)
\ee
by requiring   $F(\zo=\zt=0 |R )=0$.

\section{Evaluation of  the bulk-annulus amplitude}

The bulk correlation in the annulus in (\ref{bulk-annulus})
is calculated in $1/R_0$ expansion.
First note that $\xi$  is the function of $x$ and its explicit form
can be found by expanding  around $\xi = \tilde \xi =1$,
\begin{align}
x
&=- Q_p(u) |_{\l =0}
=-(\xi-1)  - {(\xi- 1)^2 } \left( \frac{  L'_p (1)}2 \right)
- {(\xi-1)^3} \left( \frac{ L^{(2)}_{p}(1)}6 \right) + O( (\xi-1)^4),
\nonumber
\\
\xi&=1-x-{x^2}\left( \frac{L'_{p}(1) }2 \right)
-{x^3} \left( \frac12 \left( L'_{p}(1) \right)^2\right.
-\left.\frac{1}{6}  L^{(2)}_p (1)\right) +O(x^4) \,.
\end{align}
Then  $g(x, \tilde y)$ in (\ref{g-function}) is expanded as (with $m+n=3$ )
\begin{align}
\label{ge}
g_e(x, \tilde y)
&= g_e^{(0)} + g_e^{(1)} (x + \tilde y)
+  g_e^{(2)} (x^2 + \tilde y^2 )
+ g_e^{(1,1)} x \tilde y
+O(x^m \tilde y^n)
\\
\label{go}
g_o(x, \tilde y)
&=(x -\tilde y) R_0  \left(  g_o^{(0)} + g_o^{(1)} (x + \tilde y)
+  g_o^{(2)} (x^2 + \tilde y^2 + x \tilde y  )
+O(x^m \tilde y^n) \right)
\end{align}
with  the coefficients $   g_e^{(i)}$ and $g_o^{(i)}$  which  depend on $p$ and $k$ only.
Explicit calculation shows that  $g_e^{(0)}=g_o^{(1)}=0$.
In addition, the term with $g_e^{(1)}$ vanishes when integration is done
due to the exchange symmetry of $x$ and $\tilde y$.
The term with $ g_o^{(0)}= (p+1-k)(p-k)/2 $ is $R_0$ independent 
when integrated out and the rest terms give 
\be
 \int_{x \tilde y}\frac{e^{- |x - \tilde y| R_0}} {|x- \tilde y|^2}
\Big(  g_e^{(2)} (x^2 + \tilde y^2 )
+ g_e^{(1,1)} x \tilde y
+ R_0 g_o^{(2)} (x -\tilde y) (x^2 + \tilde y^2 + x \tilde y  )
\Big)
=  \frac A { R_0^2} 
\ee
with 
$A= \frac23 g_e^{(2)} - \frac16 g_e^{(1,1)} -  g_o^{(2)} =0 $.

\section{Summation formula}
We provide useful summation formula.
When $p$ and $a$ are integers and $1\le a <p$, one has
\be
\label{sum1}
\sum_{j=1}^{p-1}
\frac{\sin^2(\pi  j a/p)}{\cosh(\pi j/p)- \cosh (\pi \xi/p) }
=- p \frac{  \sin (\pi \xi (1-a/p))\sin( \pi  \xi a/p)}
{\sinh(\pi \xi) \sinh (\pi \xi/p)} \,.
\ee
One can check that both sides have same poles and residues.
In addition, the leading behavior as $\xi \gg 1$, the leading behavior
is $e^{- \pi \xi/p}$ with the same coefficient, $-p$.

When $p$ and $a,b$ are integers and $1\le a, b <p$, one has  \cite{kutasov}
\be
\label{sum2}
\sum_{j=1}^{p-1}
\frac{\sin(\pi  j a/p) \sin (\pi j b/p)}{\cosh(\pi j/p)- \cosh (\pi \xi/p) }
=- p \frac{ \sin( \pi  \xi (1-A-B)) \sin (\pi \xi (A-B))}
{\sinh(\pi \xi) \sinh (\pi \xi/p)}
\ee
where $A=(a+b)/(2p)$ and $B=|a-b|/(2p)$.
This can be obtained from (\ref{sum1})
by changing  the numerator of LHS as two terms of sine squared
using  the formula $\sin x \sin y = \sin^2 \frac{x+y}2 - \sin^2 \frac{x-y}2$.

When $q_1<q_2$ are co-prime numbers  with  integers $k_1$ and $ k_2$
 ($1\le k_1 <q_1$ and $1\le k_2 <q_2$), one has
\be
\label{sum3}
\sum_{\a=1}^{q_1-1}
\sum_{\b=-(q_2 -1)}^{q_2-1}
\frac{\sin(\pi  t k_1 /q_1) \sin (\pi t  k_2/q_2)}{\cosh(\pi t/\t)- \cosh (\pi \xi/\t) }
= -2\t   \frac{ \sin( \pi  \xi (1-A-B)) \sin (\pi \xi (A-B))}
{\sinh(\pi \xi) \sinh (\pi \xi/\t)}
\ee
where $A=k_1/q_1 +  k_2 /q_2 $, $B=|k_1/q_1 -  k_2 /q_2|$,
$t= \a q_2 + \b q_1$ and $\t \equiv q_1q_2$.

Finally, when $m$ and $\ell$ are integers
$1 \le m, \ell < q$, one has
\be
\label{sum4}
\frac{ \sin (x m /q)}{\sin(x/q) }
\frac{ \sin (x \ell /q) }{\sin(x/q)} =
\sum _{k}  C_k ^{m, \ell} \sin( x k/q) /\sin(x/q)
\ee
where $ C_k ^{m, \ell}$ is a certain integer
and satisfies  $ C_k ^{m, \ell} =  C_k ^{q-m, q-\ell}$.


\end{document}